\documentclass[pra,aps,twocolumn,superscriptaddress,showpacs,amsmath,amssymb,floatfix]{revtex4}
\usepackage{comment}

\usepackage{graphicx}
\usepackage{epstopdf}

\begin{document}

\title{Localization in momentum space of ultracold atoms in incommensurate lattices}

\author{M. Larcher}
\affiliation{INO-CNR BEC Center and Dipartimento di Fisica, Universit\`a di Trento, 38123 Povo, Italy}
\author{M. Modugno}
\affiliation{Department of Theoretical Physics and History of Science,UPV-EHU, 48080 Bilbao, Spain}
\affiliation{IKERBASQUE, Basque Foundation for Science, 48011 Bilbao, Spain}
\author{F. Dalfovo}
\affiliation{INO-CNR BEC Center and Dipartimento di Fisica, Universit\`a di Trento, 38050 Povo, Italy}

\begin{abstract}
We characterize the disorder induced localization in momentum space for 
ultracold atoms in one-dimensional incommensurate lattices, according to 
the dual Aubry-Andr\'e model.  For low disorder the system is localized in 
momentum space, and the momentum distribution exhibits time-periodic 
oscillations of the relative intensity of its components. The 
behavior of these oscillations is explained by means of a simple three-mode 
approximation. We predict their frequency and visibility by using typical 
parameters of feasible experiments. Above the transition the system diffuses 
in momentum space, and the oscillations vanish when averaged over different 
realizations, offering a clear signature of the transition.
\end{abstract}
\pacs{03.75.Lm, 03.75.Kk}

\maketitle

\section{Introduction}

Disorder is among the most intriguing and ubiquitous aspects of condensed 
matter \cite{kramer}. The prediction of localization induced by a random 
disorder in a periodic lattice dates back to the seminal work by Anderson 
\cite{anderson1958}. Recently, ultracold atomic gases have demonstrated to 
be an extremely versatile tool to explore the effects of disorder, owing 
to the great tunability of the system parameters and geometrical 
configurations \cite{fallani2008,billy2008,roati2008,deissler2010,demarco}. 
Notably, Anderson localization of matter waves has been observed both for 
correlated (speckle) disorder \cite{billy2008} and quasi-periodic optical 
lattices \cite{roati2008}, the latter case realizing the so-called 
Aubry-Andr\'e (AA) model \cite{harper1955,aubry1980}. Localization effects 
in the quantum dynamics of one-dimensional lattice models have attracted a 
large interest also in the recent theoretical literature 
\cite{larcher2009,interaction_vs_localization,others}. The AA model has the 
peculiar property of exhibiting a transition, already in one dimension and
both in real and momentum space (duality), from extended to exponentially 
localized states as the disorder strength is increased above a critical value 
\cite{aubry1980,aulbach2004,lahini2009}.  So far the evolution of wave packets in the 
AA model has been investigated mainly in real space, looking for signatures 
of the transition from ballistic spreading to sub-diffusion and localization,  
both in theory \cite{hiramoto1988,larcher2009} and experiments 
\cite{roati2008, lahini2009}.

Here we focus on the dynamics of the momentum distribution and identify 
measurable effects of the transition from diffusion to localization in 
momentum space. This is relevant for current experiments with ultracold atoms, 
where the momentum distribution is accessible {\it via} time of flight 
measurements and, typically, with an higher accuracy than in real space. 
In addition, it provides complementary information for a better 
understanding of the key role played by duality of the AA model.

\section{Aubry-Andr\`e model}

The quantum dynamics of the AA model for the amplitude $\psi_n(t)$ is 
governed by the equation
\begin{equation}
	i \partial_t \psi_n = -[\psi_{n+1}+\psi_{n-1}]+\lambda\cos(2\pi\alpha n+\varphi)\psi_{n}
	\label{eq:AA}
\end{equation}
where $\lambda$ is the disorder strength, and $\varphi$ an arbitrary phase. 
This equation can be used to model non-interacting atoms subject to two 
periodic optical lattices with different wavelengths, that is, a bichromatic 
lattice \cite{roati2008,deissler2010}. In this case, $\alpha$ is the 
ratio between their wavelengths. In the tight-binding regime, Eq.~(\ref{eq:AA}) 
is easily obtained starting from the Schr\"odinger equation and projecting 
the continuous wave function $\psi(x)$ on the basis of the Wannier functions, 
$w_n(x)$, of the lowest band of the primary lattice \cite{modugno2009}, 
$\psi(x) = \sum_n \psi_n w_n(x)$, where $n$ is the lattice site index. 
The secondary lattice appears in the last term of Eq.~(\ref{eq:AA}) in 
the form of a modulation which mimics a disorder (quasi-disorder). Without 
any loss of generality, one can choose $\alpha < 1$, because Eq.~(\ref{eq:AA}) 
is invariant under a shift of $\alpha$ by an integer number. When $\alpha$ 
is a rational number one can write it as $\alpha=p/q$ and the solution of 
Eq.~(\ref{eq:AA}) can be restricted to a region of size $N=q$, which 
coincides with spatial periodicity of the system. The case of irrational 
$\alpha$ can be obtained as a limit of a continued fraction approximation.  

Aubry-Andr\'e showed \cite{aubry1980} that this model has a duality under 
the following transformation 
\begin{equation}
	{\phi_l}= N^{-1/2} \sum_n \psi_n e^{in[{2\pi}\alpha l+\theta]}e^{-i\varphi l}
	\label{eq:AA_transformation_phi}
\end{equation}
which maps Eq.~(\ref{eq:AA}) into an equation for the new variable $\phi_l$ 
exactly of the same form as Eq.~(\ref{eq:AA}) but with disorder strength 
$4/\lambda$.  This transformation corresponds to a projection on a basis of 
quasi-momentum eigestates with eigenvalues $\xi = 2\pi\alpha l+\theta= 
(2\pi/N) p l+\theta$, where $\xi$ is defined in the first Brillouin zone, 
$\xi\in[-\pi,\pi]$. By using this duality one can show that the AA model 
undergoes a transition from an extended to a localized regime at $\lambda=2$.

In the AA model the space is discrete and one can conveniently introduce
the discrete Fourier transform (DFT)
\begin{equation}
	f_k =  N^{-1/2} \sum_n \psi_n e^{-i\frac{2\pi}{N}kn}
	\label{eq:DFT}
\end{equation}
which corresponds to a projection on the basis of quasi-momentum 
eigenstates with eigenvalues $\xi=(2\pi/N)k$. The relation bewteen 
momentum and quasi-momentum distributions is 
$|\tilde{\psi}(k,t)|^2 =  {N}|f_k(t)|^2 |\tilde{w}(k)|^2$
where $\tilde{w}(k)$ is the Fourier transform of the Wannier function 
centered on the lattice site $n=0$. Note also that the expression 
(\ref{eq:DFT}) is related to the dual mapping (\ref{eq:AA_transformation_phi}) 
by an arbitrary shift of $\theta$ and a permutation \cite{aulbach2004}. 
By applying the transformation (\ref{eq:DFT}) to the AA model one 
gets \cite{modugno2009} 
\begin{equation}
\label{eq:AA_momentum}
i \partial_t f_k = -2\cos\left(\frac{2\pi k}{N}\right)f_k+\frac{\lambda}{2}\left[ e^{-i\varphi}f_{k+p}+e^{i\varphi}f_{k-p} \right]
\end{equation}
which is the equation describing the evolution in momentum 
space. Owing to the duality of the AA model under the transformation 
(\ref{eq:AA_transformation_phi}), and to the similarity of the latter 
with Eq.~(\ref{eq:DFT}), the localization properties in momentum space 
are the same of the dual AA model, except for the fact that disorder 
couples modes differing by $|\Delta \xi| = 2\pi\alpha$ instead of 
neighboring ones.

\begin{figure}[t!]
		\centerline{\includegraphics[width=\columnwidth]{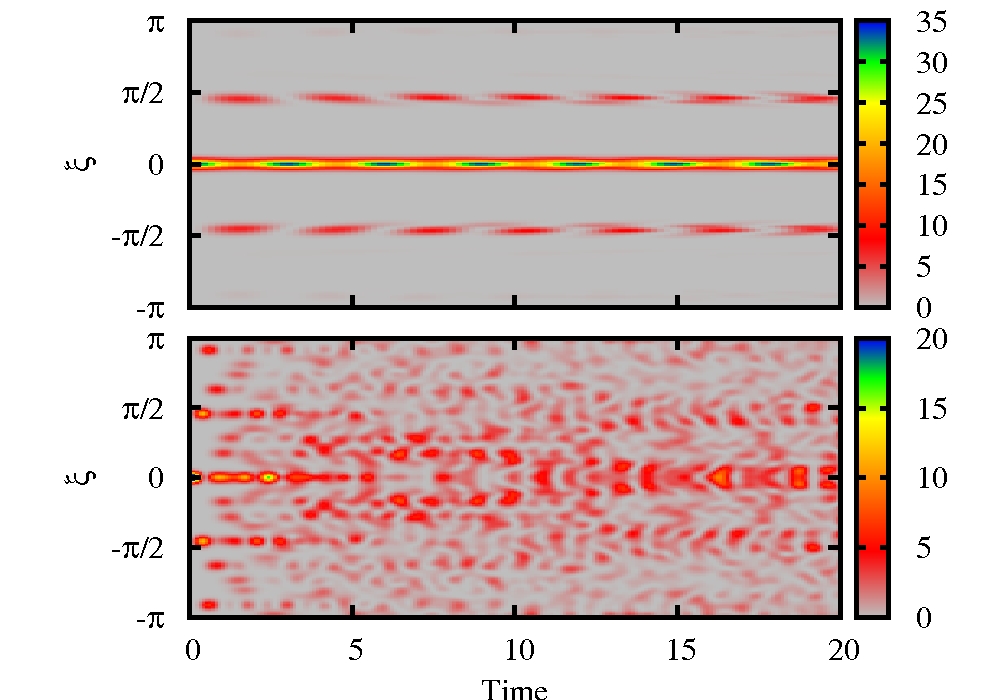}}
	\caption{(Color online) Quasi-momentum distribution $|f_k(t)|^2$ 
obtained from the DFT of the solution of  Eq~(\ref{eq:AA}). The index $\xi$ is related to 
$k$ by $\xi=2\pi k/N$.  Here we use $\alpha=0.2282...$ and $\varphi=0$. The 
initial wave packet in real space is a Gaussian of width  $\sigma=10$. Time is 
given in dimensionless units. Top panel: $\lambda=1$, only few modes are 
involved and a periodic oscillation of the central and side peaks is observed. 
The side peaks are at a distance $\pm2\pi\alpha$ from the central peak. 
Bottom panel: $\lambda=5$, the evolution is affected by the coupling of 
many modes and the periodic oscillations are no more visible. } 
	\label{fig:transition_eps}
\end{figure}

\begin{figure}[b]
	\centerline{\includegraphics[width=0.9\columnwidth]{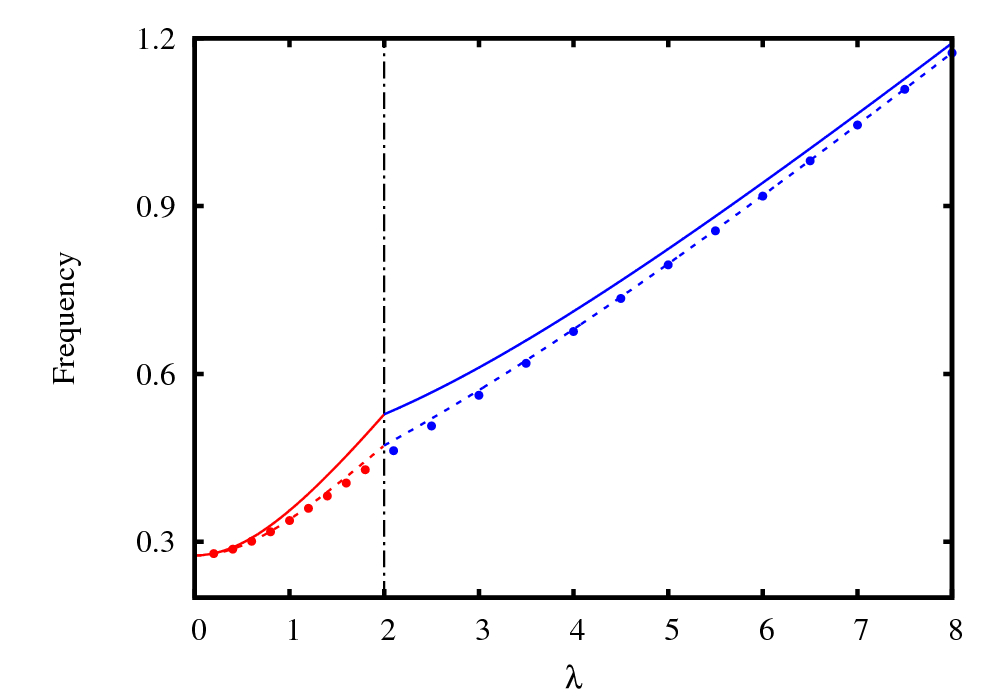}}
	\caption{(Color online) 
Oscillation frequency of the central peak of quasi-momentum ($\lambda<2$, red)  
and spatial ($\lambda>2$, blue) distributions, for $\sigma=10$,  $\varphi=0$. 
The full solution of the AA model (dots) is compared with the predictions 
of an analytic three-mode approximation (full lines) and a semi-analytic 
five-mode approximation (dashed line).}
\label{fig:period_oo}
\end{figure}

\section{Results and discussions}

In the following we will consider the evolution of a wave packet, by 
solving Eq.~(\ref{eq:AA}) in real space and then computing the evolution 
in momentum space by means of the DFT \cite{numerics}. As initial condition 
we choose a Gaussian wave packet, $\psi_j(0)=A\exp\{-{j^2}/{2\sigma^2}\}$, 
where $\sigma$ is the width and $A$ is a normalization factor. This choice 
is convenient if one wants to simulate realistic experimental configurations; 
it also allows one to explore the behavior of sharp to broad wave packets 
in a continuous manner.  

According to the previous discussion, localization in momentum space occurs 
for $\lambda<2$, where the wave packet instead spreads in real space. In 
this regime one thus expects to see only one or few momentum components 
significantly populated. Conversely, for $\lambda>2$ the regime is diffusive 
in momentum space and localized in real space, and one should see a
momentum distribution with many modes coupled together during the evolution 
of the system. This is indeed confirmed by our numerical simulations, 
as shown in Fig.~\ref{fig:transition_eps} (the value $\alpha=1064.4/866.6-1=0.2282...$ 
has been chosen in order to model the bichromatic lattice of the experiment 
of Ref.~\cite{deissler2010}). A striking feature is that, for $\lambda<2$, 
the quasi-momentum components $|f_k|^2$ exhibit periodic oscillations, 
occurring among the central peak and two side peaks at distance 
$\pm 2\pi\alpha$. The numerical result for the frequency of these oscillations 
is plotted in Fig.~\ref{fig:period_oo} as a function of $\lambda$ 
(red points for $\lambda <2$) \citep{frequency}. 

This oscillating behavior can be understood in terms of the following
model. Let us assume that the width $1/\sigma$ of the initial quasi-momentum 
distribution is small enough, so that only the $k=0$ mode can be considered 
populated at $t=0$ ($f_k(0)=\delta_{k,0}$) and the time evolution couples 
the mode at $k=0$ with two side modes at $k=\pm p$ only (i.e., $\xi=0 $ 
and $\pm2\pi\alpha$, respectively). This assumption is valid when 
$1/{\sigma}\lesssim 2\pi\alpha$. In this way, the AA equation is mapped into 
an eigenvalue problem of a $3\times 3$ matrix,  whose eigenvectors and 
eigenvalues can be written as $g_{k,j}$ and $E_j$, respectively, with 
$j=1,2,3$. The initial condition is $f_k(0)=\sum_{j=1}^3 \gamma_j g_{k,j}$, 
where the coefficients $\gamma_j$ are given by the standard rules of quantum 
mechanics. Under these assumptions one has $\gamma_{j=3}\equiv 0$, and the 
time evolution takes the form
\begin{align}
	|f_k(t)|^2=&(\gamma_1 g_{k,1})^2+(\gamma_2 g_{k,2})^2\nonumber\\
	&+\gamma_1\gamma_2 g_{k,1}g_{k,2}\cos\left[(E_2-E_1)t\right] \, .
	\label{eq:periodic_behaviour}
\end{align}
This expression describes a time-periodic oscillation of the relative 
intensity of the central and side peaks, with frequency 
$\nu(\lambda<2) =|E_2-E_1|/2\pi$, given by    
\begin{equation}
	\nu (\lambda<2) = \pi^{-1} \sqrt{[1-\cos(2\pi\alpha)]^2+\lambda^2/2 } \, .
\label{eq:momentum_period}
\end{equation}
It is worth stressing that, once $\alpha$ is fixed, this frequency depends only 
on the disorder strength $\lambda$, but not on the phase $\varphi$. This 
three-mode approximation provides a reasonable description of the numerical 
results, as shown by the solid line for $\lambda <2$ in Fig.~\ref{fig:period_oo}. 

\begin{figure}[t!]
	\centerline{\includegraphics[width=\columnwidth]{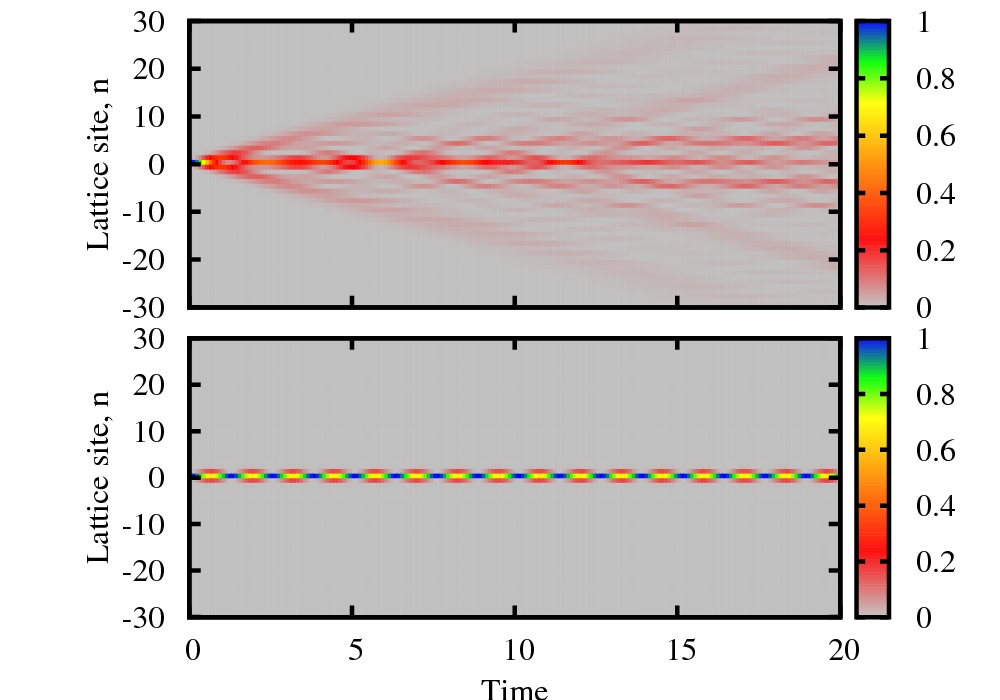}}
	\caption{(Color online) Spatial distribution $|\psi_n|^2$ obtained from the 
solution of  Eq~(\ref{eq:AA}) using a single site initial condition $\psi_n=\delta_{n,0}$.  
As in Fig.~\protect\ref{fig:transition_eps}, we use $\alpha=0.2282...$ and $\varphi=0$. 
Time is given in dimensionless units.  Top panel: $\lambda=1$, the initially localized 
wavepacket spreads ballistically and there are no visible periodic oscillations. 
Bottom panel: $\lambda=5$, the wave packet is localized and a periodic oscillation
of the central peak and its nearest neighbors is clearly visible.} 
\label{fig:transition_real_eps}
\end{figure}

The three-mode approximation becomes inaccurate when approaching $\lambda=2$, 
where more modes are coupled during the evolutions. In order to check this 
effect, one can go one step furher and consider a five-mode approximation in
which the time evolution couples the modes $k=0$, $k=\pm p$ and $k=\pm 2p$ 
(i.e., $\xi=0 $, $\pm2\pi\alpha$, and $\pm4\pi\alpha$). This is 
straightforward generalization of the three-mode approximation, except 
for the fact that the differential equations for the coefficients $\gamma_j(t)$
do not yield simple analytical expressions and, moreover, the solutions 
contain several oscillation frequencies. The red dashed line in the 
$\lambda<2$ part of Fig.~\ref{fig:period_oo} is our numerical result
for the dominant component of the frequency spectrum, solution of the 
five-mode approximation, which mostly determines the time evolution of
the central peak. As expected we get a better agreement with the full 
integration of Eq~(\ref{eq:AA}) compared to the three-mode 
approximation, especially in the region close to the transition 
point $\lambda=2$.   

\begin{figure}[t]
	\centerline{\includegraphics[width=0.9\columnwidth]{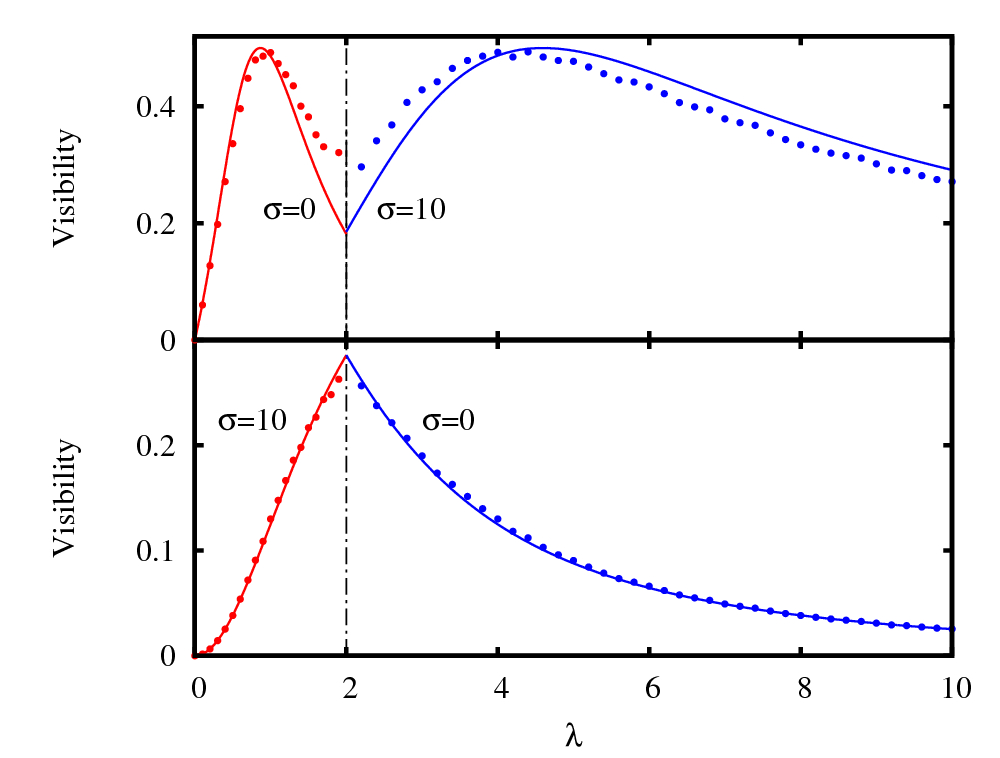}}
	\caption{(Color online) Visibility of the oscillations in real and 
quasi-momentum space a function $\lambda$, for $\varphi=0$. The numerical 
points are compared with the three-mode approximation (full lines). 
Bottom: only few modes have been initially populated by using a 
$\delta$-function initial wave packet in real space and a Gaussian 
with $\sigma=10$ in momentum space. Top: many modes have been initially 
populated by inverting the initial conditions with respect to the bottom 
panel. 
}
	\label{fig:visibility_oo}
\end{figure}

In the region $\lambda>2$ the few-mode approximation is expected to fail 
in the quasi-momentum space, where the wave packet is no more localized. 
Indeed, in this regime, we do not see any significant evidence of periodic 
behaviors in the quasi-momentum distribution (see the bottom panel of 
Fig.~\ref{fig:transition_eps}). Conversely, owing to the duality of the 
AA model, one expects periodic oscillations to take place in real space, 
where the wave packet is localized. This is confirmed by our numerical 
integration of Eq.~(\ref{eq:AA}), as shown in Fig.~\ref{fig:transition_real_eps}.
In the top panel the wave packed spreads ballistically and one 
cannot detect any significant periodic behaviour; conversely in the bottom 
panel the wave packet is localized, while the central peak and its nearest 
neighbours oscillate periodically. By assuming that the initial density 
distribution is localized in a single lattice site, $n=0$, which is coupled  
with the nearest neighboring sites, $n=\pm 1$, we obtain a three-mode 
approximation analogous to the one used before in quasi-momentum space, 
but describing oscillations in the spatial distribution. The scenario 
in real space is more complicated because one generally observes oscillations 
with several frequency components, which also depend on the phase $\varphi$. 
However, in the special case $\varphi=0$, one finds just a single 
frequency, given by 
\begin{equation}
	\nu(\lambda>2) =  (2\pi)^{-1} \sqrt{\lambda^2 [1-\cos(2\pi\alpha)]^2 + 8} \, ,
	\label{eq:real_period}
\end{equation}
which is shown as the solid line for $\lambda>2$ in Fig.~\ref{fig:period_oo}.
In the same figure we also plot the frequency obtained from the full numerical 
integration of Eq.~(\ref{eq:AA}) (blue dots in the $\lambda>2$ region); in 
this calculation we have used a Gaussian of width $\sigma=10$ as initial 
shape of the wave packet, but we have also checked that the frequency $\nu$ 
does not depend on $\sigma$, except close to $\lambda=2$. 
The dashed line is the result of a straightforward semi-analytic extension 
to five modes, as in the $\lambda<2$ region. It is worth stressing that 
the condition for the validity of the few-mode approximation for the
oscillations in real space ($\psi_n=\delta_{n,0}$ or, equivalently, 
$\sigma \lesssim 1$) is much more constraining than the one in momentum 
space ($1/{\sigma}\lesssim 2\pi\alpha$) from the point of view of 
experimental realization. 

The amplitude of the oscillations in both real and momentum space also 
changes with $\lambda$, affecting 
its visibility. The latter can be calculated from the frequency spectrum 
of the numerical solution of Eq.~(\ref{eq:AA}), as the ratio between the 
modulus of the Fourier component of frequency $\nu(\lambda)$  and the 
modulus of the component at zero frequency. In a consistent way, one can 
define the visibility in the three-mode approximation; for the oscillations 
in momentum space for $\lambda<2$, the visibility can be written as 
$(1/2)\gamma_1\gamma_2 g_{0,1}g_{0,2}/[(\gamma_1 g_{0,1})^2+(\gamma_2 g_{0,2})^2]$. 
A similar definition can be given in real space for $\lambda>2$. 
In Fig.~\ref{fig:visibility_oo} we show the visibility of the oscillations 
as a function of $\lambda$. The points are the numerical results, while 
the lines represent the three-mode approximation. We have used two values 
for the width of the initial Gaussian wave packet, namely $\sigma=0$ 
(i.e., a $\delta$-function) and $\sigma=10$. In the upper panel, the 
two values of $\sigma$ are used for $\lambda<2$ and $\lambda>2$, respectively. 
They correspond to a broad initial wave packet both in momentum space for 
$\lambda<2$ and real space for $\lambda>2$. In the bottom panel we use 
again the same values of $\sigma$, but in the opposite regions, so to have 
a narrow initial wave packet in both spaces \citep{width}. One can see 
that the visibility depends significantly on both $\sigma$ and $\lambda$. 
Again, the three-mode approximation is qualitatively correct, except near 
$\lambda=2$. We observe that the three-mode approximation gives a better 
agreement for a narrow initial distribution (lower panel), as in the 
opposite case of a broad distribution many modes are initially excited 
and this approximation is not expected to be accurate. Another interesting 
feature is the effect of the duality of the AA model. Indeed, in both panels, 
the results in the region $\lambda<2$ almost coincide with those in the 
region $\lambda>2$ under the change of variable $\lambda \to 4/\lambda$, 
provided the initial distributions are broad (upper panel) or narrow 
(lower panel) in both momentum and real spaces; this duality also implies 
the continuity at $\lambda=2$.

\begin{figure}
\centerline{\includegraphics[width=1\columnwidth]{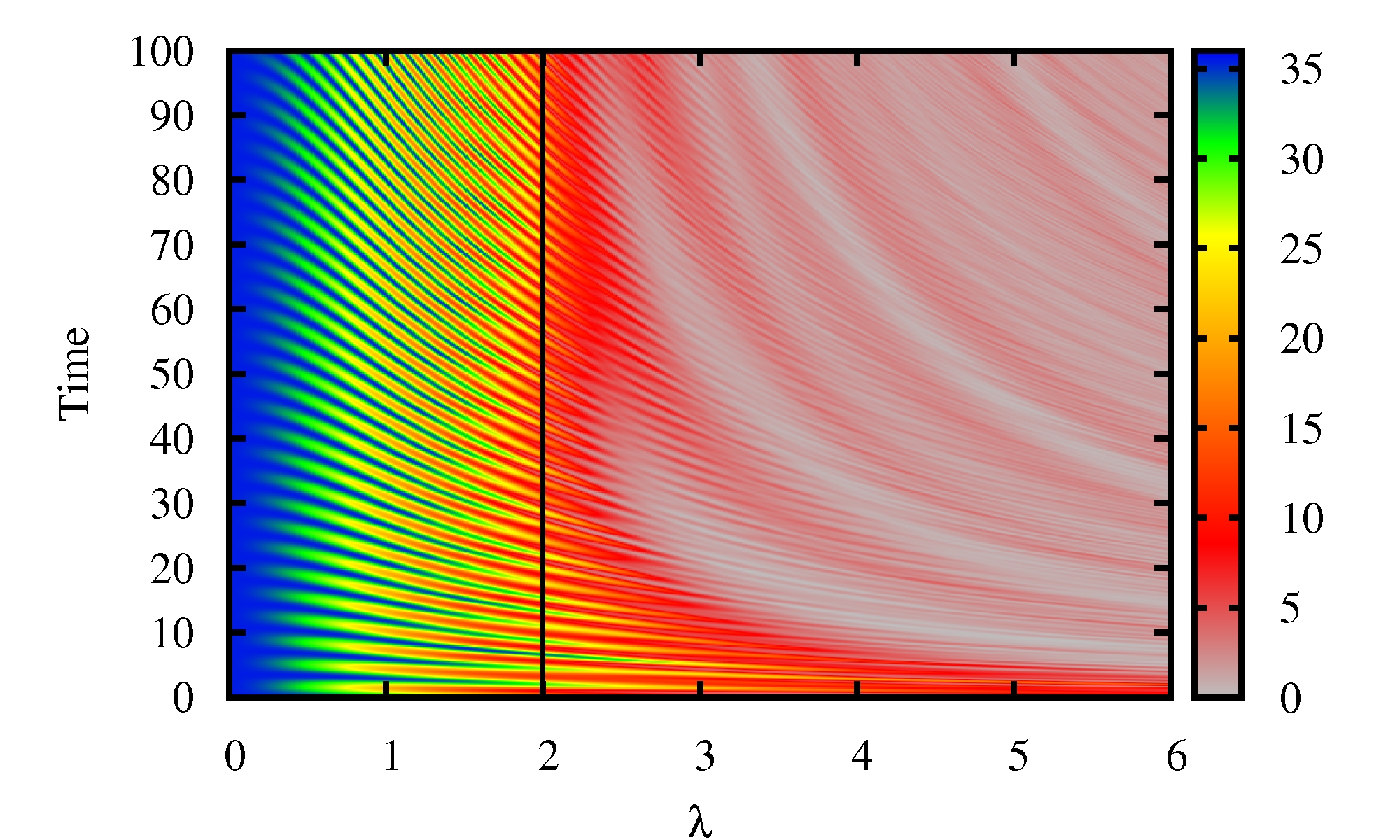}}
\caption{(Color online) Phase-averaged intensity of the central peak in 
momentum space, $|f_{0}(t)|^2$, as a function of time and disorder 
strength $\lambda$ for a wavepacket with $\sigma=10$. The intensity is 
given in arbitrary units. 
	}
	\label{fig:color_plot_central_peak}
\end{figure}

\section{Summary}

We have shown that the time evolution of a wave packet in the 
Aubry-Andr\'e model exhibits interesting periodic behaviors 
both in momentum space, for $\lambda<2$, and in real space, for 
$\lambda>2$. The occurrence of oscillations in the momentum and
density distributions can be used to test the applicability of 
the AA model to the description of the dynamics of ultracold 
gases in 1D bichromatic lattices. We have numerically calculated the frequency 
and visibility of these oscillations and we have introduced a simple few-mode 
approximation to interpret their behavior.  The use of momentum space has 
some important advantages. First, the frequency of the oscillations does 
not depend on the relative phase $\varphi$ of the two lattices or, in other 
terms, on the initial position of the wave packet, which would be hardly 
controllable in typical experiments. Second, the condition for the 
applicability of a few-mode approximation is less restrictive than in 
real space, because the width of the initial wave packet in momentum 
space can be easily made much smaller than the distance between the 
modes which are coupled in time evolution. 

Our analysis suggests that the oscillations of the central and side peaks in the 
momentum distribution can be efficiently used to probe the transition 
from diffusion to localization in the AA model. A possible strategy 
consists of measuring the intensity of the central peak 
as a function of time for different values of $\lambda$, exploiting the 
fact that for $\lambda>2$ the oscillations are phase dependent, while 
for $\lambda<2$ they are not. Actually, in typical experiments with 
ultracold atoms, the phase $\varphi$ varies at random at each realization, 
so that performing an average over many realizations at fixed $\lambda$ is 
equivalent to an average over numerical simulations with different $\varphi$. 
Thus one expect that the oscillations vanish for $\lambda>2$ (phase 
sensitive regime), but remain clearly visible for $\lambda<2$ (phase 
independent regime). This is shown in Fig.~\ref{fig:color_plot_central_peak}, 
where the average has been done over $50$ different values of the phase 
$\varphi$ for each value of $\lambda$.  Indeed the behavior of 
$|f_0(t)|^2$ exhibits a transition at $\lambda=2$. From the same figure 
one can also extract the frequency $\nu(\lambda<2)$. By using the 
experimental parameters of Ref.~\cite{deissler2010}, with 
$\alpha=1064.4/866.6$ and $\lambda=1$, the oscillation period 
turns out to be of the order of $5$~ms.   

These results are relevant for current experiments with ultracold atoms, 
where the momentum distribution can be detected with good resolution
by performing time of flight measurements. This study provides also a
starting point for future investigations on the interplay between 
interaction and localization. It is known that in certain regimes the 
presence of a repulsive interaction can destroy the disorder induced 
(Anderson) localization giving a sub-diffusive spreading of an initially 
localized wavepacket \citep{larcher2009,interaction_vs_localization}, but 
its experimental detection in real space is currently limited by the 
difficulty to reach the long time regime, and by the finite resolution 
in the observation of low density tails of the atomic distribution 
\cite{deissler2010}. In this sense, the observation of oscillations in 
momentum space can be a more reliable tool. 

{\it Acknowledgments.} M.M. acknowledges support from INO-CNR through the DQS EuroQUAM 
project from ESF. This work is also supported by Miur.


\begin{thebibliography}{}
\bibitem{frequency}
The frequency is extracted from the frequency spectrum of the central peak in quasi-momentum (real) space, $|f_0(t)|^2$ ($|\psi_0(t)|^2$). The signal is Fourier transformed, and the oscillation frequency is identified as the dominant component of the frequency spectrum.
\bibitem{width}
For $\sigma=10$, the momentum width $1/\sigma$ is much smaller than the distance between coupled modes, $\Delta\xi=2\pi\alpha$.
\end{thebibliography}
\end{document}